\documentclass[graybox]{svmult}

\usepackage{type1cm}        
\usepackage{makeidx}         
\usepackage{graphicx}        
\usepackage{multicol}        
\usepackage[bottom]{footmisc}

\usepackage{newtxtext}       %
\usepackage{newtxmath}       

\makeindex             
                       
\usepackage{relsize}
\usepackage[normalem]{ulem}
\usepackage{amsmath}

\usepackage{svg}
\usepackage{graphicx}
\usepackage{subcaption}
\usepackage{wrapfig}
\usepackage{fancyvrb}                       

\begin{document}

\title*{Sequenceable Event Recorders}
\author{Luca Cardelli}
\institute{Luca Cardelli \at University of Oxford, Oxford UK, \email{luca.a.cardelli@gmail.com}}
%
%
\maketitle

\abstract{With recent high-throughput technology we can synthesize large heterogeneous collections of DNA structures, and also read them all out precisely in a single procedure. Can we use these tools, not only to do things faster, but also to devise new techniques and algorithms? In this paper we examine some DNA algorithms that assume high-throughput synthesis and sequencing. We record the order in which $N$ events occur, using $N^2$ redundant detectors but only $N$ distinct DNA domains, and (after sequencing) reconstruct the order by transitive reduction.}

\section{Introduction}
\label{sec:1}

With recent high-throughput technology we can synthesize large heterogeneous collections of DNA structures \cite{Ellington2017, Sinyakov2021}, and also read them all out precisely in a single procedure \cite{Reuter2015}. This contrasts with the older practice of assembling structures one at a time, and of reading them out individually (e.g., by fluorescence), or reading them together ambiguously (e.g., by gel electrophoresis). Can we take advantage of these high-throughput and high precision technologies, not only to do things faster, but also to devise new techniques and algorithms? In this paper we examine some DNA algorithms that assume both high-throughput synthesis and high-throughput sequencing: they would not be very practical otherwise.

A sequence `\,s\,' of DNA nucleotides \emph{hybridizes} (forms a double strand) with its reverse Watson-Crick complement denoted `\,s*\,'; we write the resulting double strand as `\,\uline{s}\,'. Subsequences of `\,s\,' are called \emph{domains} provided they are \emph{independent} of each other, that is, provided that differently identified domains do not hybridize with each other, or with significantly long parts of each other \cite{Zhang2011}. Under normal laboratory conditions, a domain `\,a\,' is called \emph{short} if it hybridizes reversibly with `\,a*\,', and \emph{long} if it hybridizes irreversibly with it. 

A short single-stranded domain `\,t\,', called a toehold, followed in the same sequence by a long single-stranded domain `\,a\,' can initiate \emph{strand displacement}. This is the process (later detailed in Figure \ref{fig:4wayBranchMigration}) where a single-stranded sequence `\,ta\,' hybridizes to a double strand composed of `\,t*\,' attached to the bottom strand of a double stranded `\,\uline{a}\,'. The invading `\,ta\,' can displace and possibly replace the existing `\,a\,' domain of the double strand through a random-walk competition between the two `\,a\,' domains hybridizing to the same `\,a*\,'. 

A \emph{nick} is an interruption in one of the two strands of a double strand, at the boundary between two domains. By cascading short and long domains, occasionally separated by nicks, we can achieve multi-step strand displacements where the whole sequence of displacements can itself be reversible or irreversible. This way we can emulate reversible and irreversible chemical reactions \cite{Soloveichik5393} and other computational abstractions \cite{Qian2011}.

The readout of the outcome of such sequences of displacements is often done by fluorescence. Fluorophore/quencher pairs are attached to some domains that participate in the reactions, those in particular whose displacement indicates that a significant event has occurred. The displacement separates the fluorophore from the quencher and hence induces visible fluorescence. This provides a real-time account of the computation, but the readout capability is restricted: a limited number of separate events can be detected by using different fluorescence colors. This is analogous to debugging a program by inserting a limited number of print statements at a time, each one printing a single letter.

Another way of achieving a readout is via gel electrophoresis, to distinguish the sequences in a solution by their length at the end of the experiment (or at predetermined time points). Many different sequences can be identified provided they have different lengths (masses), and provided that we know their length ahead of time. Unexpected lengths can be hard to identify. This is analogous to debugging a program by using control flow counters to tell us how many times each routines is invoked or each structure is accessed, without any insight about the order of events.

Finally, and especially with more recent high-throughput technology, we can obtain a readout by ligating the nicks and sequencing all the strands in the solution at the end of the experiment (or at predetermined time points). With high-throughput sequencing we can inspect potentially the entire composition of the solution. The debugging analogy now is that of taking a core dump: analyzing in complete detail the entire state of a computation, but only infrequently or at the end, and again without any obvious insight on the order of events that occurred.

The order of events is usually of great interest: for example, multiple laborious gene knockout experiments are frequently carried out to determine the order of gene activations. What if we could instead take a single core dump that tells us the order of all the events of interest? To that end, we should record the order of events within the state of the system, so that we can inspect such recording at the end as part of the core dump. Assuming high-throughput sequencing, we can embed a large amount of information within the solution. We are going to assume that we can embed $N^2$ pieces of information, where $N$ is the number of events of interest. This seems achievable for reasonably small $N$ while providing a lot of information, encoding for each event whether it happened before, together, or after any other event. Each one of the $N^2$ event order detectors is a structure that accepts inputs but does not produce outputs: when it detects certain conditions it locks down in a stable state and waits to be sequenced later. 

Our strategy is therefore to embed a \emph{preorder} (a reflexive and transitive relation) of events within the solution. This is a \emph{pre}-order because we may not be able to detect the precise order of two events if they happen very close to each other, in which case both directions are recorded. With $N^2$ detectors we can determine the order of any pair of events without needing to coordinate the detectors with each other or with a central structure, hence each detector can be relatively simple. An alternative is to use only $N$ detectors that sequentially add records to a central \emph{tape}, but this requires a way of guaranteeing atomic access to the tape \cite{Qian2011}. Still, event recorders of the tape variety, readable by sequencing, have been nicely demonstrated using natural DNA and protein mechanisms \cite{Shipman2017, Tanna2020}.

A preorder is not the entire history of a computation. We are considering the preorder of \emph{first-occurrence} of events: any subsequent occurrences for the same signal are not recorded. This limited information can still provide support for causality: if an event always precedes another event over a number of runs, then this supports the first event causing the second, or having a common cause. 

In the rest of this paper we aim to describe the architecture of such a preorder recorder, using DNA strand displacement technology, slowly building up from simpler problems. A property of all the designs in this paper is that (apart from the single-stranded input signals) all DNA structures are nicked double strands with no additional modifications or secondary structure. Therefore, the required and potentially large numbers of components can be fabricated by bacterial cloning as a single or a few long DNA double strands, followed by enzymatic cutting and nicking \cite{Chen2013} (see Appendix). Other technologies for high-throughput synthesis of large heterogeneous libraries exist \cite{Ellington2017, Sinyakov2021}. Thus, we rely on both high-throughput synthesis for producing the $N^2$ detectors, and on high-throughput sequencing to read them out. 

\section{Occurrence Recorder}
\label{sec:2}

We begin by investigating the simplest event recorder: recording the occurrence of events at any time during an experiment. By an `event' here we mean the appearance of a whole population of identical molecules, and in fact a specific structure of molecules that can be uniformly identified. Any event that does not fit that description, must first be transduced into one of these uniform molecular structures. By a \emph{signal} we mean a population of one such molecular species over time, and by an \emph{event} we mean the appearance of a signal population (we do not detect the disappearance of a population).

In discussions we summarize DNA structures by a textual notation. In addition to lowercase letters like `\,a\,' for single-stranded long domains, and underlined letters like `\,\uline{a}\,' for the corresponding double-stranded long domains, a single short domain is used for all toeholds: `\,\textbf{--}\,' is an open (i.e., un-hybridized) toehold on a single strand or on the upper strand of a double strand, `\,\uline{\,\,\,}\,' is an open toehold on the lower strand of a double strand, and `\,\uline{\textbf{--}}\,' is a covered (double-stranded) toehold. A sequence of domains on a double strand with an initial open toehold and an intermediate covered toehold looks like `\uline{\,\,\,a\textbf{--}b}'. This summary notation omits information about nicks, which are instead detailed in corresponding figures. Note that, before sequencing, all the open domains should be complemented and all the nicks should be ligated.

Figures instead depict the corresponding single and double strands graphically (e.g., Figure \ref{fig:FluorescenceOccurrenceRecorder}). A domain is a short or long sequence of dashes `\,\verb|-|\,' with domain delimiters `\,\verb|>|\,' and `\,\verb|<|\,' pointing in the 5'-to-3' direction to indicate either a nick (an interruption in the strand) or the 3' end, and `\,\verb|+|\,' to indicate the 5' end or the logical boundary of a domain (not a nick). The name of a domain is a lower case letter placed on top of the upper strand, with implicitly the reverse complement domain on the lower strand. All toeholds are the same sequence: they have a blank name. Reversible reactions are `\verb|<=>|' and irreversible reactions are `\verb|=>|'.

\subsection{Yes Gate}
The events that we want to detect are represented by single-strands `\,\textbf{--}a\,' consisting of a (short) toehold `\,\textbf{--}\,' attached to a (long) domain `\,a\,'. If `\,\textbf{--}a\,' is ever present, we want to know about it: this is the purpose of the Yes gate for  `\,a\,'.

\begin{figure}[htbp]
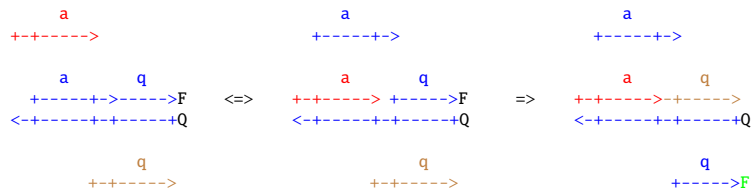

\centering
\begin{Verbatim}[fontsize=\relsize{-1}, commandchars=\\\{\}]
         \textcolor{red}{a}                            \textcolor{blue}{a}                            \textcolor{blue}{a}
    \textcolor{red}{+-+----->}                      \textcolor{blue}{+-----+->}                    \textcolor{blue}{+-----+->}
    
         \textcolor{blue}{a       q}                    \textcolor{red}{a}       \textcolor{blue}{q}                    \textcolor{red}{a}       \textcolor{brown}{q}
      \textcolor{blue}{+-----+->----->}F    <=>    \textcolor{red}{+-+----->} \textcolor{blue}{+----->}F     =>    \textcolor{red}{+-+----->}\textcolor{brown}{-+----->} 
    \textcolor{blue}{<-+-----+-+-----+}Q           \textcolor{blue}{<-+-----+-+-----+}Q           \textcolor{blue}{<-+-----+-+-----+}Q

                 \textcolor{brown}{q}                            \textcolor{brown}{q}                            \textcolor{blue}{q}
            \textcolor{brown}{+-+----->}                    \textcolor{brown}{+-+----->}                      \textcolor{blue}{+----->}\textcolor{green}{F}
\end{Verbatim}
\caption{Yes gate for detection by fluorescence.}
\label{fig:FluorescenceOccurrenceRecorder}
\end{figure}

First (Figure \ref{fig:FluorescenceOccurrenceRecorder}) let us consider the traditional way of detecting `\,\textbf{--}a\,'. A double-stranded structure `\uline{\,\,\,a\textbf{--}q}' with an open toehold `\,\uline{\,\,\,}\,' accepts the single-strand `\,\textbf{--}a\,' (reversibly) and opens up another toehold, yielding `\uline{\textbf{--}a\,\,\,q}'. That structure then locks down (irreversibly) by combining with an auxiliary single-strand `\,\textbf{--}q\,' to produce the fully hybridized `\uline{\textbf{--}a\textbf{--}q}' and the toehold-free `\,q\,'. 

If we attach a fluorophore (F) and quencher (Q) pair at the right end of `\uline{\,\,\,a\textbf{--}q}' (and not at the end of `\,\textbf{--}q\,'), we can detect the occurrence of `\,\textbf{--}a\,' because it separates F from Q and causes visible fluorescence. However, if we were to (ligate and) sequence the solution, it would be difficult or impossible to tell the difference between the initial and final state, because they differ only by open toeholds and by the positions of nicks that are erased by ligation.

\begin{figure}[htbp]
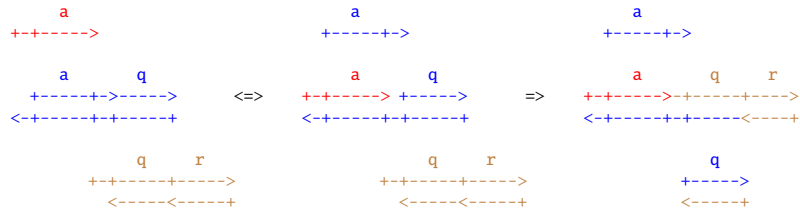

\centering
\begin{Verbatim}[fontsize=\relsize{-1}, commandchars=\\\{\}]
         \textcolor{red}{a}                             \textcolor{blue}{a}                            \textcolor{blue}{a}
    \textcolor{red}{+-+----->}                       \textcolor{blue}{+-----+->}                    \textcolor{blue}{+-----+->}
    
         \textcolor{blue}{a       q}                     \textcolor{red}{a}       \textcolor{blue}{q}                    \textcolor{red}{a}       \textcolor{brown}{q     r}
      \textcolor{blue}{+-----+->----->}      <=>    \textcolor{red}{+-+----->} \textcolor{blue}{+----->}      =>    \textcolor{red}{+-+----->}\textcolor{brown}{-+-----+---->}
    \textcolor{blue}{<-+-----+-+-----+}             \textcolor{blue}{<-+-----+-+-----+}            \textcolor{blue}{<-+-----+-+-----}\textcolor{brown}{<----+}

                 \textcolor{brown}{q     r}                       \textcolor{brown}{q     r}                      \textcolor{blue}{q}
            \textcolor{brown}{+-+-----+----->}               \textcolor{brown}{+-+-----+----->}                \textcolor{blue}{+----->}
              \textcolor{brown}{<-----<-----+}                 \textcolor{brown}{<-----<-----+}                \textcolor{brown}{<-----+}
\end{Verbatim}
\caption{Yes gate for detection by sequencing.}
\label{fig:OccurrenceRecorder}
\end{figure}

\begin{figure}[htbp]
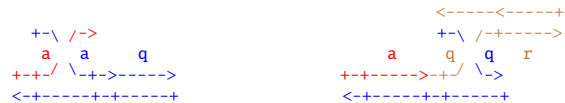

\centering
\begin{Verbatim}[fontsize=\relsize{-1}, commandchars=\\\{\}]
                                                               \,\,\textcolor{brown}{<-----<-----+}
                      \textcolor{blue}{+-\textsubscript{\textbackslash}}\,\,\,\textcolor{red}{\textsubscript{\text{\reflectbox{\textbackslash}}}->}                                   \textcolor{blue}{+-\textsubscript{\textbackslash}}\,\,\,\textcolor{brown}{\textsubscript{\text{\reflectbox{\textbackslash}}}-+----->}
                       \textcolor{red}{a}   \textcolor{blue}{a     q}           \,\,             \textcolor{red}{a}     \textcolor{brown}{q}   \textcolor{blue}{q}   \textcolor{brown}{r}
                    \textcolor{red}{+-+-\textsuperscript{\text{\reflectbox{\textbackslash}}}}\,\,\,\textcolor{blue}{\textsuperscript{\textbackslash}-+->----->}                 \textcolor{red}{+-+----->}\textcolor{brown}{-+-\textsuperscript{\text{\reflectbox{\textbackslash}}}}\,\,\,\textcolor{blue}{\textsuperscript{\textbackslash}->}
                    \textcolor{blue}{<-+-----+-+-----+}                 \textcolor{blue}{<-+-----+-+-----+}

\end{Verbatim}
\caption{3-way and 4-way displacements in the first and last steps of Figure \ref{fig:OccurrenceRecorder}.}
\label{fig:4wayBranchMigration}
\end{figure}

Let us now consider (Figure \ref{fig:OccurrenceRecorder}) an additional domain `\,r\,' that will help us tag the desired outcome. The `\,\textbf{--}q\,' single-strand is replaced by a `\,\uline{\textbf{--}qr}\,' double strand, but with a nick on the bottom between `\,q\,' and `\,r\,' \footnote{The `q' domain can be single-stranded, interacting by a simpler 3-way displacement, but that would rule out producing the structure directly by cloning \cite{Chen2013}. If the whole `\,\textbf{--}qr\,' were single stranded, a polymerase could not attach to the final structure to complement the `\,r\,' domain, as required for sequencing and positive detection, but a PCR step could be used instead.}. The first reaction is the same as before, but the second reaction is now a 4-way strand displacement \footnote{4-way strand displacement is slower than 3-way \cite{Nadine2013} Ch.5 (although potentially more robust \cite{Duose2012}). This may degrade the ability of our algorithms to separate events in time, but otherwise it does not affect their logic, which includes the possibility of coincidence of events. The 4-way displacement is of the unusual `open' kind \cite{Nadine2013}, that is, initiated by a single toehold binding instead of two.} (Figure \ref{fig:4wayBranchMigration} right). This detector is non-catalytic: it captures some of the `\,\textbf{--}a\,' strands and releases `\,a\textbf{--}\,' strands (which are usually harmless).

If this gate is triggered, then the main outcome is `\uline{\textbf{--}a\textbf{--}qr}', which is a nicked but fully complemented double strand: it is ready for ligation and sequencing. If the gate is not triggered, then the outcome is the initial `\uline{\,\,\,a\textbf{--}q}' which is distinguishable after sequencing \footnote{`\uline{\,\,\,a\textbf{--}q}' needs to be fully complemented, ligated, and sequenced. To that end, we can add an additional double-stranded domain on the left of the initial toehold, as in \cite{Seelig2021}. This allows a polymerase to proceed in the 3'-5' direction of the bottom strand and fully complement the top strand. For this presentation we omit these domains because they have no other function and do not participate in the described reactions. Moreover, even if sequencing misread the initial state, we would still get our answer by the presence or absence of the final state `\uline{\textbf{--}a\textbf{--}qr}' + `\uline{q}'.}.

\begin{figure}[htbp]
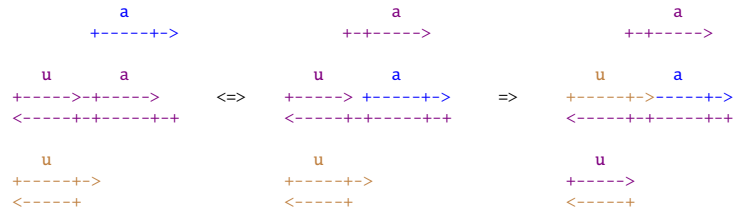

\centering
\begin{Verbatim}[fontsize=\relsize{-1}, commandchars=\\\{\}]
               \textcolor{blue}{a}                           \textcolor{violet}{a}                            \textcolor{violet}{a}    
            \textcolor{blue}{+-----+->}                 \textcolor{violet}{+-+----->}                    \textcolor{violet}{+-+----->}  
                       
       \textcolor{violet}{u       a}                   \textcolor{violet}{u}       \textcolor{blue}{a}                    \textcolor{brown}{u}       \textcolor{blue}{a}
    \textcolor{violet}{+----->-+----->}      <=>    \textcolor{violet}{+----->} \textcolor{blue}{+-----+->}     =>     \textcolor{brown}{+-----+->}\textcolor{blue}{-----+->}
    \textcolor{violet}{<-----+-+-----+-+}           \textcolor{violet}{<-----+-+-----+-+}            \textcolor{violet}{<-----+-+-----+-+}

       \textcolor{brown}{u}                           \textcolor{brown}{u}                            \textcolor{violet}{u} 
    \textcolor{brown}{+-----+->}                   \textcolor{brown}{+-----+->}                    \textcolor{violet}{+----->}
    \textcolor{brown}{<-----+}                     \textcolor{brown}{<-----+}                      \textcolor{brown}{<-----+}
\end{Verbatim}
\caption{Catalytic Yes gate, additional reactions.}
\label{fig:CatalyticOccurrenceRecorder}
\end{figure}

For a catalytic version (one that does not sequester the input), consider the design in Figure \ref{fig:CatalyticOccurrenceRecorder}: we add two more structures to Figure \ref{fig:OccurrenceRecorder} that absorb the `\,a\textbf{--}\,' that was left over and convert it back to a free `\,\textbf{--}a\,'. Such catalytic irreversible gates avoid sequestering weak signals, while being fully activated by weak signals, leading to robust detection (if the signals are not drained too quickly by downstream processing).

\subsection{Occurrence Recorder Algorithm}

We can use Yes gates to detect a collection of signals in an experiment via a single high-throughput readout: we prepare a Yes gate detector for each signal, we mix them in at the beginning, and we sequence the entire solution at the end, revealing any detectors that have fired.

\section{Coincidence Recorder}
\label{sec:3}

We now move to a more interesting task: detecting the simultaneous presence of signals. The idea enabling the sequencing-based readout of gates, and in particular the novel use of 4-way displacement, is due to Yuan-Jyue Chen and Georg Seelig \cite{Seelig2021} (the Yes gate of Figure \ref{fig:OccurrenceRecorder} is also a special case of this). Their design was originally meant as an And gate made of a sequenceable Join part accepting inputs, and a sequenceable Fork part producing outputs. We are going to use just a sequenceable Join half to detect the simultaneous occurrence of any pair of signals in a given set of signals, relying on high-throughput sequencing to inspect all possible combinations.

\subsection{Join Gate}

\begin{figure}[htbp]
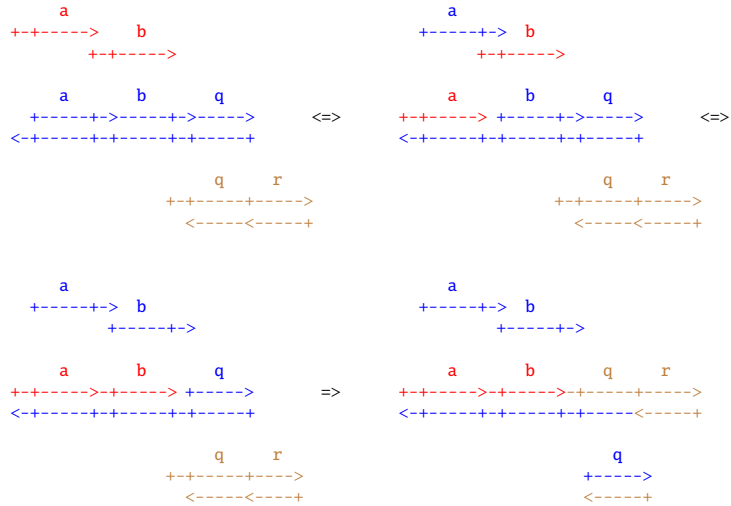

\centering
\begin{Verbatim}[fontsize=\relsize{-1}, commandchars=\\\{\}]
         \textcolor{red}{a}                                       \textcolor{blue}{a}
    \textcolor{red}{+-+----->}    \textcolor{red}{b}                            \textcolor{blue}{+-----+->}  \textcolor{red}{b}
            \textcolor{red}{+-+----->}                               \textcolor{red}{+-+----->}

         \textcolor{blue}{a       b       q}                       \textcolor{red}{a}       \textcolor{blue}{b       q}
      \textcolor{blue}{+-----+->-----+->----->}      <=>      \textcolor{red}{+-+----->} \textcolor{blue}{+-----+->----->}      <=>     
    \textcolor{blue}{<-+-----+-+-----+-+-----+}               \textcolor{blue}{<-+-----+-+-----+-+-----+}

                         \textcolor{brown}{q     r}                                 \textcolor{brown}{q     r}
                    \textcolor{brown}{+-+-----+----->}                         \textcolor{brown}{+-+-----+----->}
                      \textcolor{brown}{<-----<-----+}                           \textcolor{brown}{<-----<-----+}


         \textcolor{blue}{a}                                       \textcolor{blue}{a}
      \textcolor{blue}{+-----+->  b}                            \textcolor{blue}{+-----+->  b}
              \textcolor{blue}{+-----+->}                               \textcolor{blue}{+-----+->}

         \textcolor{red}{a       b}       \textcolor{blue}{q}                       \textcolor{red}{a       b}       \textcolor{brown}{q     r}
    \textcolor{red}{+-+----->-+----->} \textcolor{blue}{+----->}       =>      \textcolor{red}{+-+----->-+----->}\textcolor{brown}{-+-----+----->}
    \textcolor{blue}{<-+-----+-+-----+-+-----+}               \textcolor{blue}{<-+-----+-+-----+-+-----}\textcolor{brown}{<-----+}

                         \textcolor{brown}{q     r}                                  \textcolor{blue}{q}
                    \textcolor{brown}{+-+-----+---->}                             \textcolor{blue}{+----->}
                      \textcolor{brown}{<-----<----+}                             \textcolor{brown}{<-----+}
\end{Verbatim}
\caption{Join gate for detection by sequencing.}
\label{fig:CoincidenceRecorder}
\end{figure}

The design in Figure \ref{fig:CoincidenceRecorder} is rooted in a fluorophore-oriented Join gate, along the lines of Figure \ref{fig:FluorescenceOccurrenceRecorder}, which ultimately comes from \cite{Seelig2006} and \cite{Cardelli2013}. However, again here we want to find a sequencing-friendly version, where the initial structures `\uline{\,\,\,a\textbf{--}b\textbf{--}q}' and `\,\textbf{--}\uline{qr}\,' with input signals `\,\textbf{--}a\,' and `\,\textbf{--}b\,' are sequencing-distinguishable from the final structure `\uline{\textbf{--}a\textbf{--}b\textbf{--}qr}', which indicates that both signals were present at the same time. The gate locks down when the two signals are received in turn. If one signal appears first and persists until the second arrives, this gives the same result as both signal appearing together. If one signal is removed before the other one appears, the gate reverts and the result indicates no co-occurrence. The gate can be made more kinetically symmetrical by mixing Join(a,b) with Join(b,a).

\begin{figure}[htbp]
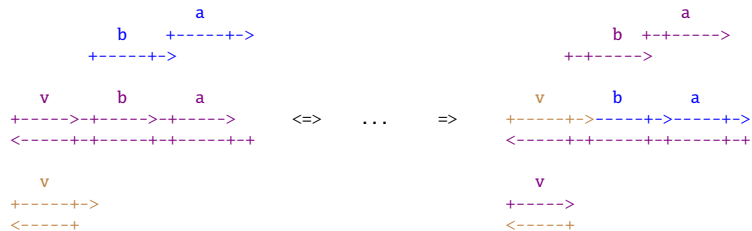

\centering
\begin{Verbatim}[fontsize=\relsize{-1}, commandchars=\\\{\}]
                       \textcolor{blue}{a}                                                 \textcolor{violet}{a}
               \textcolor{blue}{b    +-----+->}                                     \textcolor{violet}{b  +-+----->}
            \textcolor{blue}{+-----+->}                                        \textcolor{violet}{+-+----->}
                       
       \textcolor{violet}{v       b       a}                                  \textcolor{brown}{v}       \textcolor{blue}{b       a}
    \textcolor{violet}{+----->-+----->-+----->}      <=>    ...     =>     \textcolor{brown}{+-----+->}\textcolor{blue}{-----+->-----+->}
    \textcolor{violet}{<-----+-+-----+-+-----+-+}                          \textcolor{violet}{<-----+-+-----+-+-----+-+}

       \textcolor{brown}{v}                                                  \textcolor{violet}{v}
    \textcolor{brown}{+-----+->}                                          \textcolor{violet}{+----->} 
    \textcolor{brown}{<-----+}                                            \textcolor{brown}{<-----+} 
\end{Verbatim}
\caption{Catalytic Join gate, additional reactions.}
\label{fig:CatalyticCoincidenceRecorder}
\end{figure}

As in the previous case, we can add structures to this gate that convert it to a catalytic gate. But we need to handle the two signals together in the additional structures, because `\,a\textbf{--}\,' must be able to revert to `\,\textbf{--}a\,' when `\,\textbf{--}b\,' is not present. Hence we use the binary structure in Figure \ref{fig:CatalyticCoincidenceRecorder} for distinct a,b. This structure cannot coexist with a catalytic Yes(a) as it would lock down Join(a,b) on the first input: a later non-coincident `\,\textbf{--}b\,' would give a false positive. Join(a,a) must not have the additional catalytic structures for the same reason: it is best to replace it with a non-catalytic Yes(a).

\subsection{Coincidence Recorder Algorithm}

We can use Join gates to detect the simultaneous occurrence of any pair of distinct signals in a collection: we prepare a Join gate detector for each such pair, we mix them in at the beginning, and we sequence the entire solution at the end, revealing any detectors that have fired. If we detect Join(a,b) and Join(b,c) we can deduce the coincidence of a and c, and we should also detect Join(a,c): that redundancy serves as a crosscheck. We could use fewer Join gates, but if we did not include the transitive Join(a,c) and b never came, we would not detect the coincidence of a and c.

\section{Preorder Recorder}
\label{sec:4}

We now aim to build a device to record the order of occurrence of events in an experiment. The question is: given a set of events $a,b,c,d,...$ that occur in some order, in what order did they \emph{first} occur? If some events can occur together (up to experimental uncertainty), the relationship is a \emph{preorder}: a reflexive and transitive relation. We want to reconstruct the temporal preorder of events from a single observation at the end of a run, with a single mass sequencing. 
 
Such a \emph{preorder recorder} would be useful for monitoring a process over time without sampling the system at multiple time points.  Our recorder does not record timing, does not record sequence, but it records the first-occurrence preorder, storing it within the system itself. Recording the order, rather than the full timing of events, means that we need not use energy during periods of inactivity, and we need not worry about how often we should sample the system. The energy expenditure is all preloaded: no additional resources are needed no matter how long or complex the events history becomes, and there can be no ``memory overflow'' of the recording. Repeated preorder experiments can build up evidence for causality, by observing which events always happen in the same order, independently of timing and other conditions.
 
The algorithm below uses a number of gates that is quadratic in the number of signals $N$, but is independent of the observation time. After the initial setup it requires no further energy because it reacts to signals, and does not actively inspect the environment for their presence. More subtly, the algorithm uses a number of distinct domains that is just $N+4$ (+1 for toeholds). This is important to avoid crosstalk among domains, which becomes more difficult to avoid when we have more domains. The situation would be much worse if we needed $N^2$ distinct domains in addition to $N^2$ gates.

The coincidence recorder in the previous section was obtained by iterating a Join gate. For the preorder recorder we iterate a Choice gate, which we describe next. Instead of presenting directly the domain structures (for which there are multiple possibilities), we first describe abstractly how the choice gate behaves, and how the algorithm uses it. The DNA implementation is described later.

\subsection{Choice Gate Specification}
	 
	A Choice gate is a two-input gate denoted $a?b$ between input events $a$ and $b$. As an abstract operator it is symmetric: $a?b = b?a$. Its desired behavior is as follows:
	
		\begin{itemize}
            \item If $a$ arrives no later than $b$, then $a?b$ produces a distinct result that we indicate $a\leq b$ or equivalently $b\geq a$.
            \item If $b$ arrives no later than $a$, then $a?b$ produces a distinct result that we indicate $b\leq a$ or equivalently $a\geq b$.
            \item If $a$ and $b$ arrive together, then $a?b$ produces a result that we indicate $a\sim b$ or equivalently $b\sim a$. (This is in practice an equal mixture of $a\leq b$ and $b\leq a$, or an unequal mixture if they arrive slightly offset.)
            \item As a special case, if $a$ ever arrives, then $a?a$ produces a result $a\sim a$.
        \end{itemize}
	 
	The three results between different $a$ and $b$ are assumed to be distinct and distinguishable by sequencing. Our algorithm requires only that there are three detectable final configurations: $a\leq b$ and $b\leq a$ depending on which of two inputs arrives first, and a mixture of the two, $a\sim b$, if they arrive together. We may further analyze the results quantitatively: a 100\%/0\% mixture of $a\leq b$ and $b\leq a$ indicates that enough of $a$ arrived to exhaust the gate population before any of $b$ (if any) arrived. Other mixtures may indicate how much events overlapped in time, their relative strength, or some confusion between those. Weak signals may appear to have arrived together.
	
	There are many ways to achieve this specification, and we will discuss at least two. But first we describe the algorithm that uses these gates.
	 
\subsection{Preorder Recorder Algorithm}
	 
	Suppose we have a (moderately large) set of events \emph{a,b,c,d,e,f...} , like the occurrence of some mRNAs in a cell-free extract. They will activate in some order like \emph{b.cd.ae.d} (\emph{b} first, then \emph{cd} together, then \emph{ae} together, then \emph{d}). We want to store that order as the events arise, and read it back at the end. 
	 
	For $N$ signals we need $N^2$ distinct DNA structures: all the possible combinations of two signals, including all the $x?x$ cases. We are not going to distinguish event sequences with repetitions and oscillations: we only look at the first occurrence of a signal. For example, the sequence \emph{b.b.b} is the same as \emph{b} for us, and \emph{a.b.a} is the same as \emph{a.b} (we can still tell that the first $a$ arrived before the first \emph{b}: the second \emph{a} does not confound it).
	 
	We do not provide any external timing: there are no clocks needed to sample these signals over time, and there is no predetermined sampling frequency. We just need to assume that the sequence of events is slow enough. If it is not slow enough then \emph{a.b} will look just like \emph{ab} (in practice, the closeness of two signals will be reflected in the relative proportions of \emph{a}$\leq$\emph{b} and \emph{b}$\leq$\emph{a}, so we can still get some more information). The time resolution is thus determined by the speed of the DNA reactions. If they happen to be fast enough for the intended observed system, then sampling over longer time periods does not require any more gates or any more energy: the gates just naturally sit waiting for the signals to arrive. 
	 
	 
	The input to our algorithm is a preorder of signals, like \emph{a.bc.def.g} that is occurring in real time in our experiment. We initially add to the solution all the choice gates $x?y$ such that $x$ and $y$ range over all those signals (including $x=y$). At the end we sequence all the leftover structures (e.g. $x\leq y$) and we reconstruct the preorder from them. The process of reconstructing the preorder graph from what is essentially its reachability matrix is called \emph{transitive reduction} and has the same complexity as transitive closure and matrix multiplication \cite{Aho1972}.
	 

\subsection{Crosstalking Choice Gate}
	 
We now describe a DNA implementation of the choice gate $a?b$. We discuss below how the gates crosstalk, and what are the consequences of crosstalking. But in summary, for our application this implementation is sufficient, and it is also considerably more economical than a `proper' non-crosstalking implementation.
	 
	The inputs are the usual two-domain signals with toehold on the left. For each abstract choice operator $a?b$, we use two pairs of double strands abbreviated as group $[a?b|$ and group $|b?a]$, with $a?b$ = $[a?b|$ + $|b?a]$. They are symmetric but different because $[a?b|$ reacts to a `\,\textbf{--}b\,' strand, while $|b?a]$ reacts to an `\,\textbf{--}a\,' strand. Conversely, $[a?b|$ reacts also to an `\,a\textbf{--}\,' strand and $|b?a]$ reacts also to a `\,b\textbf{--}\,' strand, through the same toehold but in opposite directions. 
	
\begin{figure}[htbp]
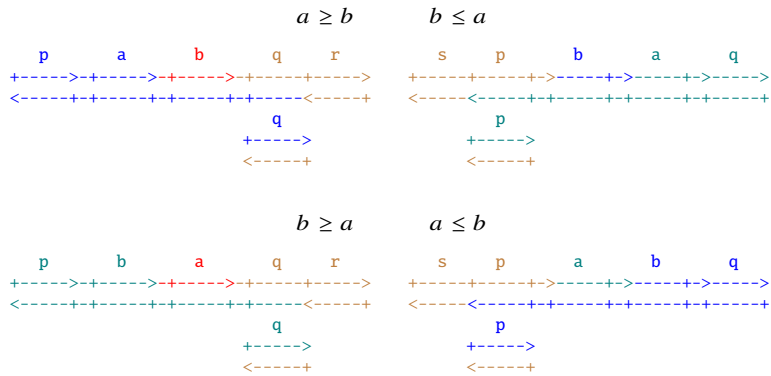
 
\centering
$a\geq b$  \,\,\,\,\,\,\,\,\,\,\,\,\,\,\,\,\, $b\leq a$
\begin{Verbatim}[fontsize=\relsize{-1}, commandchars=\\\{\}]
         \textcolor{blue}{p       a}       \textcolor{red}{b}       \textcolor{brown}{q     r}          \textcolor{brown}{s     p}       \textcolor{blue}{b}       \textcolor{teal}{a       q}
      \textcolor{blue}{+----->-+----->}\textcolor{red}{-+----->}\textcolor{brown}{-+-----+----->}    \textcolor{brown}{+-----+-----+->}\textcolor{blue}{-----+->}\textcolor{teal}{-----+->----->}
      \textcolor{blue}{<-----+-+-----+-+-----+-+-----}\textcolor{brown}{<-----+}    \textcolor{brown}{<-----}\textcolor{teal}{<-----+-+-----+-+-----+-+-----+}
                                 \textcolor{blue}{q}                      \textcolor{teal}{p}
                              \textcolor{blue}{+----->}                \textcolor{teal}{+----->}
                              \textcolor{brown}{<-----+}                \textcolor{brown}{<-----+}
                              
\end{Verbatim}
$b\geq a$  \,\,\,\,\,\,\,\,\,\,\,\,\,\,\,\,\, $a\leq b$
\begin{Verbatim}[fontsize=\relsize{-1}, commandchars=\\\{\}]                  
         \textcolor{teal}{p       b}       \textcolor{red}{a}       \textcolor{brown}{q     r}          \textcolor{brown}{s     p}       \textcolor{teal}{a}       \textcolor{blue}{b       q}
      \textcolor{teal}{+----->-+----->}\textcolor{red}{-+----->}\textcolor{brown}{-+-----+----->}    \textcolor{brown}{+-----+-----+->}\textcolor{teal}{-----+->}\textcolor{blue}{-----+->----->}
      \textcolor{teal}{<-----+-+-----+-+-----+-+-----}\textcolor{brown}{<-----+}    \textcolor{brown}{<-----}\textcolor{blue}{<-----+-+-----+-+-----+-+-----+}
                                 \textcolor{teal}{q}                      \textcolor{blue}{p}
                              \textcolor{teal}{+----->}                \textcolor{blue}{+----->}
                              \textcolor{brown}{<-----+}                \textcolor{brown}{<-----+}                    
\end{Verbatim}
\caption{Crosstalking Choice gate outcomes for $a?b$. Top: for input `\,\textbf{--}b\,' (red), which also releases back `\,\textbf{--}b\,' (teal, not shown). Bottom: for input `\,\textbf{--}a\,' (red), which also releases back `\,\textbf{--}a\,' (blue, not shown).}
\label{fig:CrosstalkingChoiceGateResult}
\end{figure}

	In Figure \ref{fig:CrosstalkingChoiceGate} each of the primary structures (top) eventually binds to one and only one of the two end caps (bottom): we arbitrarily associate one end cap with $[a?b|$ and the other with $|b?a]$ (the square bracket indicates the side the end cap is with), so in fact $a?b$ = $[a?b|$ + $|b?a]$ = $[b?a|$ + $|a?b]$ = $b?a$. The central portions with the `a' and `b' domains are surrounded by four fixed domains `s',`p',`q',`r': these are the same sequences for all the choice gates, regardless of variations in `a' and `b' \footnote{In fact, all four `s',`p',`q',`r' domains can be the same sequence without ambiguity in the outcomes of Figure \ref{fig:CrosstalkingChoiceGateResult}. Still, we keep them distinct in light of other possible constraints, such as in Figure \ref{fig:Nicking}.}. The nameless toehold is the same sequence everywhere.
\begin{figure}[htbp]
\centering
$[a?b|$ \,\,\,\,\,\,\,\,\,\,\,\,\,\,\,\,\,\,\,\,\,\,\,\,\,\,\,\,\,\,\,\,\,\,\,\,\,\, $|b?a]$
\begin{Verbatim}[fontsize=\relsize{-1}, commandchars=\\\{\}]
            \textcolor{blue}{p       a       b       q}                  \textcolor{teal}{p       b       a       q}
         \textcolor{blue}{+----->-+-----> +-----+->----->}            \textcolor{teal}{+----->-+-----> +-----+->----->}
         \textcolor{blue}{<-----+-+-----+-+-----+-+-----+}            \textcolor{teal}{<-----+-+-----+-+-----+-+-----+}
	
      \textcolor{brown}{s     p}                                                                  \textcolor{brown}{q     r}
   \textcolor{brown}{+-----+-----+->}                                                        \textcolor{brown}{+-+-----+----->}
   \textcolor{brown}{<-----<-----+}                                                            \textcolor{brown}{<-----<-----+}
\end{Verbatim}
	
\caption{Crosstalking Choice gate.}
\label{fig:CrosstalkingChoiceGate}
\end{figure}

	If a signal `\,\textbf{--}b\,' (with toehold on the left) binds to $[a?b|$, it blocks the toehold and displaces to the right. It also releases `\,b\textbf{--}\,' (with toehold to the right), which goes to $|b?a]$, again blocks the toehold there, and displaces to the left, catalytically releasing a copy of the original `\,\textbf{--}b\,'. The end caps can bind to the remaining open toeholds and lock down the configuration. If `\,\textbf{--}a\,' arrives later, it finds all the toeholds blocked and cannot bind to the remaining structures, Thus `\,\textbf{--}b\,' arriving first prevents `\,\textbf{--}a\,' from binding later. If `\,\textbf{--}a\,' arrives first, the situation is symmetric, with the end caps binding to the opposite structures than in the `\,\textbf{--}b\,'-first case.
	 
	In more detail, the initial binding of signals opens up new toeholds for the double-stranded `\,\uline{sp}\textbf{--}\,',`\,\textbf{--}\uline{qr}\,' end caps: they cause 4-way strand displacements and stabilize the outcomes in a way that is distinguishable by sequencing. For a `\,\textbf{--}b\,' input the final structures are `\,\uline{p\textbf{--}a\textbf{--}b\textbf{--}qr}\,' + `\,\uline{q}\,', which is the result we earlier called $a\geq b$, and `\,\uline{sp\textbf{--}b\textbf{--}a\textbf{--}q}\,' + `\,\uline{p}\,', which is the result we earlier called $b\leq a$ (Figure \ref{fig:CrosstalkingChoiceGateResult}, top). The opposite happens if `\,\textbf{--}a\,' arrives first (Figure \ref{fig:CrosstalkingChoiceGateResult}, bottom). If `\,\textbf{--}a\,' and `\,\textbf{--}b\,' arrive together, then both results are produced because the released `\,a\textbf{--}\,' and `\,b\textbf{--}\,' bind concurrently to as yet untouched copies of the gates.

	These activations are irreversible and catalytic: `\,\textbf{--}a\,' and `\,\textbf{--}b\,' are released back without requiring additional structures. This is going to help kinetically, and is also less likely to perturb the system we are observing. Reflexive gates $a?a$ work as expected: we need them to signify that a signal `\,\textbf{--}a\,' has arrived at some time. We produce the $a?a$ structures by the general recipe, meaning as $[a?a|$ + $|a?a]$, hence with \emph{twice} the concentration of the main structure. This is in fact what we need to keep the kinetics balanced with respect to non-reflexive gates.
	
	A single choice gate works as described, but we need to consider the situation where there are multiple choice gates together. In a gate with $[a?b|$, the input `\,\textbf{--}b\,' releases `\,b\textbf{--}\,', which goes on to bind to $|b?a]$, but also to any other $|b?x]$: crosstalk! Normally this would be incorrect, but here we want to activate $|b?x]$ as well, since it tells us that `\,\textbf{--}b\,' arrived before `\,\textbf{--}x\,'. If there is a $|b?x]$, then there is also an $[x?b|$, which driven by `\,\textbf{--}b\,' activates $|b?x]$ anyway. So the crosstalk between gates does not hurt in this particular instance. The most interesting consequence is that, as we noted, although we have $N^2$ gates, we only have to encode $N$ distinct domains (plus the 4 auxiliary ones). This greatly reduces the potential interference between domains that would be an obstacle to scaling up the number of signals. As an added benefit, these crosstalking gates are automatically catalytic (\emph{cf.} Figure \ref{fig:NonCrosstalkingChoiceGate}).
		 
	As an example, for 3 signals $a,b,c$, we use the following 9 choice gates (first column) and corresponding initial structures (second column):
\begin{align*}
	\text{ga}&\text{tes}       &          \text{struc}&\text{tures}       &  \text{after}&\text{ `\,\textbf{--}c\,'}                &  \text{after}&\text{ `\,\textbf{--}b\,'}                \\
	a&?a         &       [a?a| &\,\,\, |a?a]  &  [a?a| &\,\,\, |a?a]      &  [a?a| &\,\,\, |a?a]       \\
	b&?b         &       [b?b| &\,\,\, |b?b]  &  [b?b| &\,\,\, |b?b]      &  b\geq b &\,\,\, b\leq b   \\
	c&?c         &       [c?c| &\,\,\, |c?c]  &  c\geq c &\,\,\, c\leq c  &  c\geq c &\,\,\, c\leq c   \\
	a&?b         &       [a?b| &\,\,\, |b?a]  &  [a?b| &\,\,\, |b?a]      &  a\geq b &\,\,\, b\leq a   \\
	a&?c         &       [a?c| &\,\,\, |c?a]  &  a\geq c &\,\,\, c\leq a  &  a\geq c &\,\,\, c\leq a   \\
	b&?c         &       [b?c| &\,\,\, |c?b]  &  b\geq c &\,\,\, c\leq b  &  b\geq c &\,\,\, c\leq b
\end{align*}

	If a signal `\,\textbf{--}c\,' arrives, it initially activates 3 structures, the ones of the form $[x?c|$, producing outcomes $x\geq c$. Soon after, the signal `\,c\textbf{--}\,' that is released by those activations crosstalks with the structures of the form $|c?y]$, producing outcomes $c\leq y$ (third column). If a signal `\,\textbf{--}b\,' arrives next, it further activates some gates, but not the ones that have been used up by `\,\textbf{--}c\,' (fourth column).
	If we sequence the structures at this point, we can conclude (with multiple redundancies) that:
\begin{align*}
		c\leq b \leq a  
\end{align*}
	 
	That’s a definite $c<b$, because we observe $c\leq b$ but not $b\leq c$.
	Moreover, we do not observe  $a\leq a$ which means that $a$ never arrived. 
	If we were to observe $c\leq b$ and $b\leq c$, then we would deduce that $c,b$ arrived together, up to our time resolution.
	 
	Detection of the preorder should be robust because of the redundancies. Background noise and bad gates can be tolerated, because we just need to detect which of $a \leq b$ vs. $b\leq a$ is strongest. Moreover, our set of observed structures must be transitively closed: if the input is the sequence a.b.c then we should observe $a \leq b$ (and not $b \leq a$) and $b \leq c$ (and not $c \leq b$), and transitively also $a \leq c$ (and not $c \leq a$). The transitive closures can act as consistency checks.

\subsection{A ``Proper'' Choice Gate}
	 
\begin{figure}[htbp]
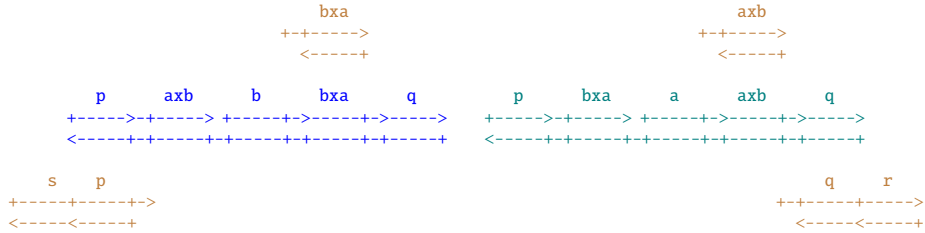

\centering
\begin{Verbatim}[fontsize=\relsize{-1}, commandchars=\\\{\}]
                                \textcolor{brown}{bxa}                                        \textcolor{brown}{axb}
                            \textcolor{brown}{+-+----->}                                  \textcolor{brown}{+-+----->}
                              \textcolor{brown}{<-----+}                                    \textcolor{brown}{<-----+}
                                
         \textcolor{blue}{p      axb      b      bxa      q}          \textcolor{teal}{p      bxa      a      axb      q}
      \textcolor{blue}{+----->-+-----> +-----+->-----+->----->}    \textcolor{teal}{+----->-+-----> +-----+->-----+->----->}
      \textcolor{blue}{<-----+-+-----+-+-----+-+-----+-+-----+}    \textcolor{teal}{<-----+-+-----+-+-----+-+-----+-+-----+}
	
    \textcolor{brown}{s    p}                                                                          \textcolor{brown}{q     r}
\textcolor{brown}{+-----+-----+->}                                                                \textcolor{brown}{+-+-----+----->}
\textcolor{brown}{<-----<-----+}                                                                    \textcolor{brown}{<-----<-----+}
\end{Verbatim}
\caption{Non-Crosstalking Choice gate.}
\label{fig:NonCrosstalkingChoiceGate}
\end{figure}

	If we want to use a choice gate in some general and modular way within some bigger design, then we need a gate that respects all the conventions, and in particular that does not crosstalk with unrelated gates. In the design in Figure \ref{fig:NonCrosstalkingChoiceGate} the domains called `\,axb\,' and `\,bxa\,' are uniquely determined by `\,a\,' and `\,b\,' to avoid crosstalk with other gates. Here a `\,b\textbf{--}\,' input does not release a `\,\textbf{--}b\,' signal that connects with other gates, but rather a `\,\textbf{--}bxa\,' signal that binds uniquely to the other half of that choice gate. In our preorder recorder application, where we use $N^2$ gates, we would now need $N+N^2$ distinct signal domains. Other than that, this choice gate could replace the crosstalking one. A catalytic version can be obtained as in Figure \ref{fig:CatalyticOccurrenceRecorder}.

\section{Conclusions}
\label{sec:5}
We have described a class of DNA algorithms designed to take advantage of high-throughput sequencing, and also relying on high-throughput synthesis. A combinatorial number of different structures are activated on demand without any timing or synchronization, operating by natural parallelism. The outcome is produced not as an output, but as the final state of the system to be read by sequencing.

\begin{acknowledgement}
Thanks to Matthew Lakin, Georg Seelig, and David Soloveichik, for helpful comments, and to Yuan-Jyue Chen and Georg Seelig for initial discussions that lead to this paper.
\end{acknowledgement}
\section*{Appendix: Restriction Enzymes}
\addcontentsline{toc}{section}{Appendix}

Naturally occurring \emph{restriction enzymes} bind to double stranded DNA and make a double-strand cut. Some natural enzymes and some engineered versions of restriction enzymes cut only one of the strands: these are called \emph{nicking enzymes} \cite{Chan2010}.
Although there are many such enzymes used for cutting natural and synthetic DNA, their properties are severely restricted\footnote{https://www.aatbio.com/data-sets/restriction-enzymes-cut-sites-reference-table}.
In nature there is also a Cas9 protein that is essentially a programmable restriction enzyme: it can cut DNA at almost any desired location determined by a separate RNA strand.

Let us imagine that we are able to design our own restriction enzymes. We show that it is then possible to cut and nick the crosstalking choice gates in Figure \ref{fig:CrosstalkingChoiceGate} out of a longer DNA double strand. Such a strand is ideally obtained by bacterial cloning, which can produce large quantities of very long high quality strands, enabling the mass production of DNA gates by cutting them out of long cloned strands \cite{Chen2013}.

\begin{figure}[htbp]
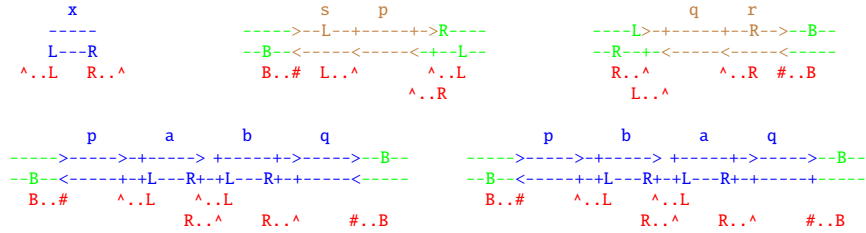

\centering
\begin{Verbatim}[fontsize=\relsize{-1}, commandchars=\\\{\}]
       \textcolor{blue}{x}                    \textcolor{brown}{     s     p}                               \textcolor{brown}{q     r}
    \textcolor{blue}{ ----- }              \textcolor{green}{-----}\textcolor{brown}{>--L--+-----+->}\textcolor{green}{R----}           \textcolor{green}{----L}\textcolor{brown}{>-+-----+--R-->}\textcolor{green}{--B--}
    \textcolor{blue}{ L---R }              \textcolor{green}{--B--}\textcolor{brown}{<-----<-----<}\textcolor{green}{-+--L--}           \textcolor{green}{--R--+-}\textcolor{brown}{<-----<-----<}\textcolor{green}{-----}      
  \textcolor{red}{^..L   R..^}            \textcolor{red}{  B..#  L..^       ^..L}               \textcolor{red}{R..^       ^..R  #..B}
                                          \textcolor{red}{^..R}                   \textcolor{red}{L..^}      
                                          
         \textcolor{blue}{p       a       b       q}                      \textcolor{blue}{p       b       a      q}
 \textcolor{green}{-----}\textcolor{blue}{>----->-+-----> +-----+->----->}\textcolor{green}{--B--}      \textcolor{green}{-----}\textcolor{blue}{>----->-+-----> +-----+->----->}\textcolor{green}{--B--}
 \textcolor{green}{--B--}\textcolor{blue}{<-----+-+L---R+-+L---R+-+-----<}\textcolor{green}{-----}      \textcolor{green}{--B--}\textcolor{blue}{<-----+-+L---R+-+L---R+-+-----+}\textcolor{green}{-----}
   \textcolor{red}{B..#     ^..L    ^..L}                          \textcolor{red}{B..#     ^..L    ^..L}
                   \textcolor{red}{R..^    R..^     #..B}                          \textcolor{red}{R..^    R..^     #..B}
\end{Verbatim}
\caption{\smaller{Restriction enzyme scheme for the crosstalking choice gate. Enzymes are shown below the DNA structures, aligned to their binding sequences, with cutting points indicated by `\texttt{\^}' (opposite strand nick) and `\texttt{\#}' (blunt double cut). Top left: common pattern for all signal domains, with the bottom strand pointing left. Top middle  and right: patterns for the end caps in context. Bottom: patterns for the main structures in context; note how the top left pattern leads to opening up the central toehold. Note also that `\texttt{L}' on a left-pointing strand cuts to the left, but on a right-pointing strand cuts to the right; similarly for `\texttt{R}' and `\texttt{B}'.}}
\label{fig:Nicking}
\end{figure}	

There are many alternatives to the hypothetical choices of restriction enzymes show below, depending on where the DNA cuts are located with respect to the binding sites of an enzyme. But let's assume the following possibilities:
\begin{itemize}
\item `\verb|B..#|' is a blunt-cutting enzyme that binds to a recognition sequence `\verb|B|' and makes a blunt double cut (i.e., at the same location on both strands) indicated by \verb|#|, at some non-critical distance towards 5'. 
\item `\verb|^..L|' is a nicking enzyme that binds to a recognition sequence `\verb|L|', and makes a nick indicated by \verb|^| towards 3', on the opposite strand, at a distance corresponding to a toehold length. 
\item  `\verb|R..^|' is a nicking enzyme that binds to a recognition sequence `\verb|R|', and makes a nick indicated by \verb|^| towards 5', on the opposite strand, at a distance corresponding to a toehold length. 
\end{itemize}


The embedding of the `\verb|B|' binding sequence is straightforward because it can be placed outside the gates, in the surrounding DNA.
The main gate structures however have several internal nicks, hence the enzyme binding sites must be placed inside the signal domains (we cannot assume they can cut precisely at a very long distance from their binding site). Since these domains occur twice with different surrounding nicks, the placement of the binding sequences is non trivial. However, the following scheme is adequate: each domain used for encoding signals has `\verb|L|' and `\verb|R|' enzyme binding sequences embedded as in Figure \ref{fig:Nicking} top left: they produce nicks at a toehold length just outside of the domain. For the staggered cutting of the end caps, though, it is non obvious that `\verb|L|' and `\verb|R|' would work together to produce a staggered double strand cut as indicated. Alternatively, two separate staggered-double-strand-cut enzymes need to be used there, with the stagger being the length of a toehold.

This scheme came from a discussion with Yuan-Jyue Chen, after he pointed out that the placement of restriction binding sequences was problematic.

\bibliographystyle{splncs04}
\bibliography{references}

\begin{thebibliography}{10}
\providecommand{\url}[1]{\texttt{#1}}
\providecommand{\urlprefix}{URL }
\providecommand{\doi}[1]{https://doi.org/#1}

\bibitem{Aho1972}
Aho, A.V., Garey, M.R., Ullman, J.D.: The transitive reduction of a directed
  graph. SIAM Journal on Computing  \textbf{1}(2),  131--137 (1972).
  \doi{10.1137/0201008}

\bibitem{Cardelli2013}
Cardelli, L.: Two-domain {DNA} strand displacement. Mathematical Structures in
  Computer Science  \textbf{23}(2),  247–271 (2013).
  \doi{10.1017/S0960129512000102}

\bibitem{Chan2010}
Chan, S.H., Stoddard, B.L., Xu, S.y.: {Natural and engineered nicking
  endonucleases—from cleavage mechanism to engineering of
  strand-specificity}. Nucleic Acids Research  \textbf{39}(1),  1--18 (08
  2010). \doi{10.1093/nar/gkq742}, \url{https://doi.org/10.1093/nar/gkq742}

\bibitem{Chen2013}
Chen, Y.J., Dalchau, N., Srinivas, N., Phillips, A., Cardelli, L., Soloveichik,
  D., Seelig, G.: Programmable chemical controllers made from {DNA}. Nature
  \textbf{8(10)},  755--762 (10 2013). \doi{10.1038/nnano.2013.189}

\bibitem{Seelig2021}
Chen, Y.J., Seelig, G.: Scaling up {DNA} computing with array-based synthesis
  and high-throughput sequencing. In this volume  (2021)

\bibitem{Duose2012}
Duose, D.Y., Schweller, R.M., Zimak, J., Rogers, A.R., N., H.W., Diehl, M.R.:
  Configuring robust {DNA} strand displacement reactions for in situ molecular
  analyses. Nucleic Acids Res.  \textbf{40(7)},  3289--3298 (2012).
  \doi{10.1093/nar/gkr1209}

\bibitem{Ellington2017}
Hughes, R.A., Ellington, A.D.: Synthetic {DNA} synthesis and assembly: Putting
  the synthetic in synthetic biology. Cold Spring Harb Perspect Biol
  \textbf{9(1)},  a023812 (1 2017). \doi{10.1101/cshperspect.a023812}

\bibitem{Nadine2013}
Nadine, D.: Synthetic molecular machines for active self-assembly: Prototype
  algorithms, designs, and experimental study. Dissertation (Ph.D.)  (2013).
  \doi{10.7907/T0ZG-PA07}

\bibitem{Qian2011}
Qian, L., Soloveichik, D., Winfree, E.: Efficient turing-universal computation
  with {DNA} polymers. In: Sakakibara, Y., Mi, Y. (eds.) DNA Computing and
  Molecular Programming. pp. 123--140. Springer Berlin Heidelberg, Berlin,
  Heidelberg (2011)

\bibitem{Reuter2015}
Reuter, J.A., Spacek, D.V., Snyder, M.P.: High-throughput sequencing
  technologies. Molecular Cell  \textbf{58(4)},  586--97 (5 2015).
  \doi{10.1016/j.molcel.2015.05.004}

\bibitem{Seelig2006}
Seelig, G., Soloveichik, D., Zhang, D.Y., Winfree, E.: Enzyme-free nucleic acid
  logic circuits. Science  \textbf{314}(5805),  1585--1588 (2006).
  \doi{10.1126/science.1132493}

\bibitem{Shipman2017}
Shipman, S.L., Nivala, J., Macklis, J.D., Church, G.M.: {CRISPR–Cas} encoding
  of a digital movie into the genomes of a population of living bacteria.
  Nature  \textbf{7663},  345--349 (07 2017). \doi{10.1038/nature23017}

\bibitem{Sinyakov2021}
Sinyakov, A.N., Ryabinin, V.A., Kostina, E.: Application of array-based
  oligonucleotides for synthesis of genetic designs. Molecular Biology
  \textbf{55},  487–500 (2021). \doi{10.1134/S0026893321030109}

\bibitem{Soloveichik5393}
Soloveichik, D., Seelig, G., Winfree, E.: {DNA} as a universal substrate for
  chemical kinetics. Proceedings of the National Academy of Sciences
  \textbf{107}(12),  5393--5398 (2010). \doi{10.1073/pnas.0909380107}

\bibitem{Tanna2020}
Tanna, T., Schmidt, F., Cherepkova, M.Y., Okoniewski, M., Platt, R.J.:
  Recording transcriptional histories using {R}ecord-seq. Nature Protocols
  \textbf{15},  513–539 (2020). \doi{10.1038/s41596-019-0253-4}

\bibitem{Zhang2011}
Zhang, D.Y.: Towards domain-based sequence design for {DNA} strand displacement
  reactions. In: Sakakibara, Y., Mi, Y. (eds.) DNA Computing and Molecular
  Programming. pp. 162--175. Springer Berlin Heidelberg, Berlin, Heidelberg
  (2011)

\end{thebibliography}

\end{document}